\documentclass[preprint]{aastex}

\newcommand{\twcott}{$^{12}$CO $J$=3$-$2}
\newcommand{\twcoto}{$^{12}$CO $J$=2$-$1}
\newcommand{\twcooz}{$^{12}$CO $J$=1$-$0}
\newcommand{\twco}{$^{12}$CO}
\newcommand{\thcott}{$^{13}$CO $J$=3$-$2}
\newcommand{\thcoto}{$^{13}$CO $J$=2$-$1}
\newcommand{\thcooz}{$^{13}$CO $J$=1$-$0}
\newcommand{\thco}{$^{13}$CO}
\newcommand{\ceott}{C$^{18}$O $J$=3$-$2}

\newcommand{\ceo}{C$^{18}$O}
\newcommand{\jft}{$J$=4$-$3}
\newcommand{\jtt}{$J$=3$-$2}
\newcommand{\jto}{$J$=2$-$1}
\newcommand{\joz}{$J$=1$-$0}
\newcommand{\etal}{et al.}
\newcommand{\eg}{e.g.}

\newcommand{\kms}{km s$^{-1}$}

\newcommand{\tmb}{$T_{\rm MB}$}

\shorttitle{Line Ratio Gradients in M82}
\shortauthors{Petitpas \& Wilson}

\begin{document}

\title{Temperature and Density Gradients Across the Nucleus of M82}
\author{Glen R. Petitpas \& Christine D. Wilson}

\affil{McMaster University, 1280 Main Street West, Hamilton Ontario,
Canada L8S 4M1}
\email{petitpa@physics.mcmaster.ca \\ wilson@physics.mcmaster.ca}

\begin{abstract}

This paper presents \twcott, \thcott, \ceott, and \thcoto\ spectra of
the irregular starburst galaxy M82 taken with the 15 m James Clerk
Maxwell Telescope. All maps exhibit a double peaked morphology, despite
evidence that the higher $J$ transitions are optically thick. This
morphology suggests that the double peaked structure is not the result
of an edge-on torus of molecular gas as is commonly assumed. We observe
line ratio gradients that can best be explained by a temperature
gradient increasing from NE to SW in conjunction with a density
gradient that increases in the opposite sense. These gradients may have
been caused by the interaction with M81, resulting in increased star
formation that both heats and depletes the molecular gas in the SW lobe
of M82.

\end{abstract}


\section{Introduction}

M82 (NGC 3034, Arp 337) is considered the prototypical starburst
galaxy. Its close proximity ($D$=3.25 Mpc, \citet{tam68}) and strong CO
lines \citep{ric75} make it an excellent location to study the effects
of a burst of star formation on molecular gas properties. Its edge-on
orientation make essential the use of optically thin molecular gas
tracers for probing the physical conditions in the interstellar
medium. Interferometric maps (\eg\ \citet{she95,nei98}) show that M82
contains three bright lobes: two main lobes on the outer edges, which
are thought to be a torus of molecular gas seen edge-on
(\eg\ \citet{til91}), and a central peak which may be housing an AGN
\citep{mux94,sea97}. Early CO studies showed that M82 had an unusually
large \twcoto/\joz\ line ratio ($\gtrsim$2, \citet{kna80,sut83}), which
suggested that the \twco\ gas in M82 is optically thin. As calibration
techniques and telescope surface quality improved, the value for the
\twcoto/\joz\ line ratio has decreased to $\sim$1 \citep{til91,wil92},
which suggests that \twco\ is likely optically thick. A high optical
depth leads to some problems in the interpretation of the double lobes
as a molecular torus. \citet{nei98} point out that if the gas were
optically thin, a torus seen from the edge would have a double peaked
appearance, while if the torus were optically thick, we should see a
elongated, bar-like structure.

Mapping a galaxy in many transitions will help our understanding of the
molecular gas dynamics since different transitions and molecules do not
necessarily trace the same gas \citep{pet98b,koh99}. While M82 has been
painstakingly mapped and studied by many authors in the lower-$J$
transitions of CO and its isotopomers (see references above),
improvements in receiver sensitivity and telescope surface quality have
made fully sampled maps of the higher-$J$ transitions of rarer CO
isotopes more tractable. This paper presents first results from \twco,
\thco, \ceott\ and \thcoto\ maps of M82. In \S\ref{obs} we describe the
observations and discuss the appearance of the spectra in comparison to
each other and previously published data. In \S\ref{pc} we use
integrated intensity line ratios to derive the physical conditions
inside the nucleus of M82 and discuss the results and possible causes
for the observed gradients.

\section{Observations and Data Reduction \label{obs}}

One of the difficulties in the interpretation of line ratios is that
the different lines are often measured with different telescopes. This
method introduces calibration uncertainties (see for example, Figure 6
of \citet{wil92}) that can be reduced if all the data are taken with
the same telescope (which is often impossible, given the large
variation in frequency between some transitions).

A variety of CO \jtt\ observations of M82 were taken using the James
Clerk Maxwell Telescope (JCMT) over the period of 1996 April 5 -- 10.
The \twcott\ map covers an area of 120\arcsec\ $\times$ 100\arcsec, the
\thcott\ map covers 56\arcsec\ $\times$ 35\arcsec, and the \ceott\ map
covers a 34\arcsec\ strip along the bar major axis. We also include
\twco\ and \thcoto\ data taken during CANSERV at the JCMT over 1993
April 22 -- 25. The data were reduced using the data reduction package
SPECX. The raw data had a linear baseline removed and then the
\jtt\ data were binned to 12.5 MHz (10.8 \kms) and the \jto\ data to 5
MHz (6.7 \kms). Since the CO \jto\ data have a different beam size and
grid spacing than the CO \jtt\ data, it was necessary to interpolate
and convolve the spectra before measuring the integrated intensity line
ratios. The data were exported to the data reduction package COMB, and
the CO \jtt\ data were convolved to the same beam size and interpolated
to the same location as the nearest CO \jto\ spectra.

The calibration was monitored by frequently observing both planets and
spectral line calibrators. The spectral line calibrators for the CO
data had intensities that were within $\sim$15\% of the published
values. We therefore adopt the published values for $\eta_{\rm MB}$ of
0.58 for the CO \jtt\ data and 0.67 for the \twcoto\ data. Pointing was
checked frequently and was determined to be accurate to within
2\arcsec\ r.m.s.

The individual \twcott, \thcott, \ceott, \twcoto, and \thcoto\ spectra
for the three bright regions of M82 are shown in Figure \ref{spec}.
The individual spectra for M82 appear asymmetric on either side of the
center, indicating rotation. The line profiles appear similar for each
transition which indicates that, at least on large scales, the emission
is coming from co-moving material. There are indications of double
peaked line profiles at the (0,0) position in the \jto\ spectra. Since
we do not see this structure as strongly in the \jtt\ spectra, it is
likely the result of the larger beam overlapping both the NE and SW
peaks simultaneously.

\section{Physical Conditions from Line Ratios \label{pc}}

It is well known that the filling factor of molecular clouds in a
galaxy is often low. Thus, integrated intensities are generally not a
good measure of the physical conditions of individual molecular clouds
because the signal is diluted by the large radio telescope beams, which
also cover much empty space. We must therefore use integrated intensity
line ratios to determine the physical conditions of individual clouds.
Using the ratio of two integrated intensities will help cancel out the
effects of beam dilution, assuming that similar regions of space are
responsible for the emission at both frequencies. The physical
conditions recovered from the analysis of line ratios are the average
conditions for all clouds within the beam.

The differing morphology in the \thco\ and \twcott\ lines results in a
gradient of the \twco/\thcott\ line ratios, increasing from east to
west (Figure \ref{ratiofig}). If both lines are optically thin, this
observation indicates a \thco/\twco\ abundance gradient across the disk
of M82. This same `abundance'-type gradient is not visible in the lower
$J$ transitions (see \twco/\thcooz\ and \jto\ ratios;
\citet{nei98,loi90}), which suggests that the \twco\ emission is not
optically thin and the gradient may be the result of a variation in
optical depth. Since optical depth is a function of density and
temperature, we have performed a Large Velocity Gradient (LVG) analysis
using a code written by Lee Mundy and implemented as part of the Miriad
data reduction package. A single component model provided a very nice
fit to the observed line ratios. We kept the \ceo\ abundance as a free
parameter, so a survey of parameter space was performed using $T_{\rm
kin}$ = 50 $\pm$ 20 K \citep{wil92}, [\twco]/[\thco] = 50 $\pm$ 20
\citep{til91}, $n$(H$_2$) from 10 to 10$^6$ cm$^{-3}$, and $N$(CO)/$dv$
from 10$^{15}$ to 10$^{20}$ cm$^{-2}$/\kms (see \citet{pet98a} for more
details). The $\pm$1$\sigma$ error bars for the line ratios shown in
Figure \ref{ratiofig} were entered into the program and the solutions
are shown in Table \ref{lvgsol}. The \twcoto/\joz\ ratios from
\citet{til91} also overlap our solutions.

In the simplest scenario, the observed line ratio gradients can be
explained by varying either temperature or density across the galaxy
while holding the other variable constant. Our analysis suggests that
either the density varies from $1.6\times10^{4}$ to $0.6\times10^{4}$
cm$^{-2}$ (for $T_{\rm kin}$ = 50 K), or the temperature varies from 70
to 30 K (for $n$(H$_2$) = 10$^4$ cm$^{-2}$) from east to west. Of
course, the gradients can also be produced by a combination of varying
$T$ and $n$, which suggests we need to include observations at other
frequencies to constrain the models.

A similar line ratio gradient was seen in HCO$^+$ \jft/\joz\ by
\citet{sea98}, but since their data combined single dish and
interferometry data, they dismissed the gradient, attributing it to an
abundance of diffuse emission on the eastern side of M82, which would
cause the HCO$^+$ emission to be underestimated. However, our model
fits show a higher density and/or temperature in the NE lobe, which
could produce an increased HCO$^+$ \jft/\joz\ ratio. Previous
\twco\ and \thcooz\ interferometric studies by \citet{kik98} find
\twco/\thcooz\ ratios that increase across M82, but in the opposite
direction from our \jtt\ ratios. This result helps rule out the
possibility that the observed gradients are the result of a \thco\ and
\ceo\ abundance gradient across M82. The \joz\ line ratio gradients are
likely the result of the higher temperatures and/or densities in the NE
lobe of M82, which bump the CO into the higher $J$ transitions, leaving
the lower $J$ levels less populated.

We can use the 850 $\mu$m continuum maps of M82 \citep{alt99} to remove
the degeneracy between density and temperature in our models. Since the
SW lobe is brighter in the 850 $\mu$m continuum maps, which trace the
product of the temperature and the column density, we can conclude that
either the column density and/or temperature is higher in the SW lobe.
In this scenario, the HCO$^+$ maps of \cite{sea98} would require that
density be higher in the NE. The pair of LVG solutions which best
satisfy the combined CO, HCO$^+$, and continuum data are flagged with
check marks in Table \ref{lvgsol}. Utilizing these additional data
sets, we therefore conclude that the observed line ratios are the
result of a temperature gradient that increases from NE to SW in
conjunction with a density gradient that increases from SW to NE.

One possible reason for this density and temperature gradient may be
interaction with the nearby galaxy M81. Gravitational interaction may
be triggering vigorous star formation in the SW lobe, which could be
heating the gas (increasing $T$) at the same time that it is being
consumed (decreasing the average density by turning the densest regions
into stars). While previous galaxy interaction simulations do not show
any indication of density gradients across the smaller galaxy, it is
unlikely that the current simulations would be able to detect a factor
of three difference in the mean density of molecular clouds across the
galaxy.

There is some dispute over the nature of the double peaked structure
seen in M82. Most studies suggest that it is the result of a molecular
torus seen edge-on. This would only be the case if the molecular gas in
the clouds were optically thin \citep{nei98}, or there were not enough
molecular clouds in the telescope beam to `shadow' the more distant
clouds. Our LVG analysis indicates that \twcooz\ transition is on the
border between optically thin and optically thick ($\tau_{J=1-0} \sim
0.1 - 3.5$). However, the \jto\ and \jtt\ maps also indicate a double
peaked structure even though these transitions are likely optically
thick (Table \ref{lvgsol}). This result can only be consistent with the
torus model if the filling factor of the optically thick clouds within
the beam is low enough that shadowing of one cloud by another is not a
problem. We can estimate the filling factor by comparing the predicted
and observed line temperatures. The LVG models predict temperatures of
19 K for \twcoto\ while our data show temperatures of $\sim$ 6 K. In
addition, the high resolution CO maps of \citet{nei98} suggest that M82
is only $\sim$ 10\arcsec\ thick and thus the filling factor of the CO
emission is $\sim$ 60\% of the 22\arcsec\ JCMT beam. This filling
factor is high enough that cloud shadowing may be important in this
region. If cloud shadowing plays a large role, the double peaked
structure in M82 could not be explained as the result of an edge-on
torus of molecular gas. Instead, it could be produced by the
accumulation of molecular gas at the Inner Lindblad Resonance radius as
it flows inward along the bar (\eg\ \citet{ken92}). There is evidence
for these double peaks (on similar size scales) in nearly face-on
barred galaxies \citep{ken92,pet98b}, where we would not expect them to
be caused by torus of gas, since the torus would have to be out of the
plane of the galaxy.

\acknowledgments We wish to thank Lorne Avery and the JCMT staff for
taking the \twco\ and \thcoto\ data during service observing. The JCMT
is operated by the Royal Observatories on behalf of the Particle
Physics and Astronomy Research Council of the United Kingdom, the
Netherlands Organization for Scientific Research, and the National
Research Council of Canada. This research has been supported by a
research grant to C.~D.~W.~from NSERC (Canada).

\newpage

\begin{deluxetable}{rccccccc}
\tabletypesize{\scriptsize}
\tablecaption{M82 LVG Solutions \label{lvgsol}}
\tablehead{\colhead{$T_{\rm kin}$}    &
           \colhead{$^{13}$X}         &
           \colhead{$^{18}$X}         &
           \colhead{log $N$(CO)/$dv$} &
           \colhead{log $n$(H$_2$)}   &
           \colhead{$\tau_{J=1-0}$}   &
           \colhead{$\tau_{J=3-2}$}   &
           \colhead{preferred solution}   \\
           } 
\startdata
\cutinhead{NE Lobe (+12,0)}
30 & 30 & 100     & 16.5 & 4.4 & 0.5 &  3  & $\surd$ \\
30 & 50 & 100     & 16.6 & 4.4 & 1.1 &  5  & $\surd$ \\
30 & 70 & 200     & 17.0 & 4.5 & 3.5 & 15  & $\surd$ \\
50 & 30 & 100     & 16.8 & 4.2 & 0.4 &  4  &         \\
50 & 50 & 150     & 17.0 & 4.2 & 1.0 &  7  &         \\
50 & 70 & 200     & 17.2 & 4.2 & 2.1 & 12  &         \\
70 & 30 & 100     & 16.7 & 4.0 & 0.1 &  3  &         \\
70 & 50 & 150-200 & 17.0 & 4.0 & 0.5 &  5  &         \\
70 & 70 & 200-250 & 17.4 & 4.0 & 1.2 &  9  &         \\
\cutinhead{SW Lobe ($-$12,0)}
30 & 30 &  75     & 16.3 & 4.0 & 0.6 &  3  &         \\
30 & 50 & 100-150 & 16.6 & 4.1 & 1.3 &  6  &         \\
30 & 70 & 200     & 16.7 & 4.1 & 1.2 &  6  &         \\
50 & 30 &  75     & 16.4 & 3.7 & 0.2 &  3  &         \\
50 & 50 & 150     & 16.8 & 3.8 & 0.6 &  5  &         \\
50 & 70 & 200     & 17.0 & 3.8 & 1.2 &  8  &         \\
70 & 30 &  75     & 16.4 & 3.6 & 0.1 &  3  & $\surd$ \\
70 & 50 & 100-150 & 16.8 & 3.7 & 0.4 &  5  & $\surd$ \\
70 & 70 & 200     & 17.0 & 3.7 & 0.9 &  8  & $\surd$ \\
\enddata 
\tablecomments{This table shows all the possible solutions for the
JCMT line ratios for $T_{\rm kin} = 50 \pm 20$ and [\twco]/[\thco] = 50
$\pm$ 20. The first column is the kinetic temperature, the second
column is the \twco/\thco\ abundance ratio, and the third column is the fit
to the \twco/\ceo\ abundance ratio. The fourth and fifth columns are the CO
column density (per \kms) and H$_2$ density required to produce the
observed line ratios, while the sixth and seventh columns are the
optical depth for the \twcooz\ and \twcott\ transitions respectively.
The last column flags the solutions that are preferred when considering
data at other frequencies (see text). }

\end{deluxetable}

\newpage

\figcaption[spectra.ps]{Individual spectra for the starburst galaxy
M82. The transitions are labeled in the same vertical order as they
appear in the plots. The (0,0) position is centered on $\alpha
=$ 09\arcdeg 51\arcmin 43\farcs 00 , $\delta =$ 069$^{\rm h}$55$^{\rm
m}$00\fs00 (B1950.0). The orientation is such that north corresponds to
an angle of 70$\arcdeg$ from the positive $x$-axis, and the bar runs
horizontally. The temperature scale is main-beam temperature. The upper
panel shows the spectra taken with (or convolved to) a 22\arcsec\ beam,
while the lower panel shows data taken with a 14\arcsec\ beam.
\label{spec}
}

\figcaption[all4ratios.ps]{Integrated intensity line ratios for
\twco/\thcoto, \twco/\thcott, \twcott/\jto, and \twco/\ceott (from top
to bottom respectively). The ratios are in main beam temperature scale
(\tmb). The uncertainties are based on the rms noise in the data,
except for the \twcott/\jto\ ratio, which is given as 20\% to reflect
the calibration uncertainties. The \jtt\ line ratios use data taken
with a 14\arcsec\ beam size; the \jto\ line ratio uses
22\arcsec\ resolution data. The \twcott\ data were convolved to a
22\arcsec\ beam to create the \twcott/\jto\ line ratios.
\label{ratiofig}
}

\plotone{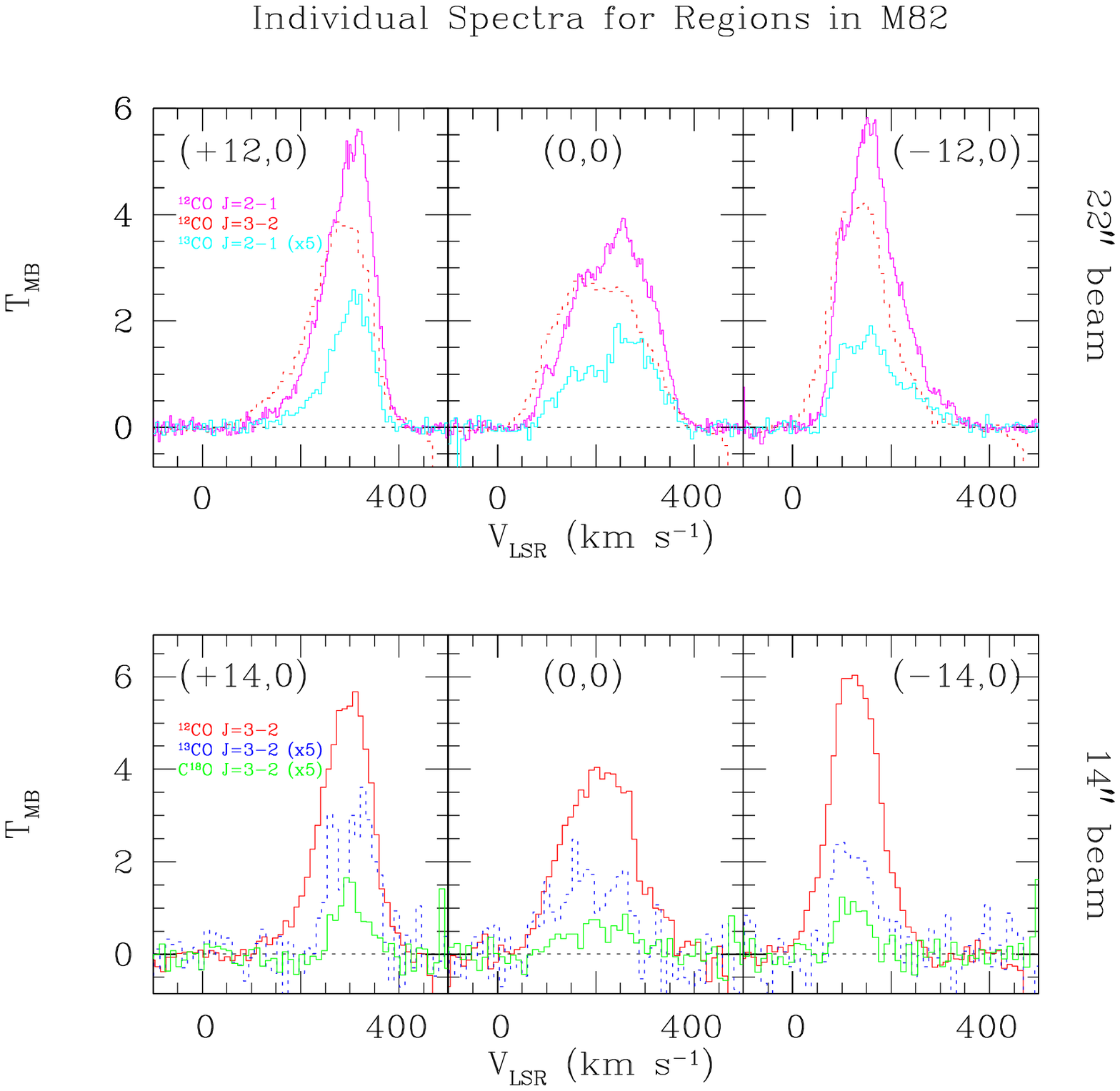}

\plotone{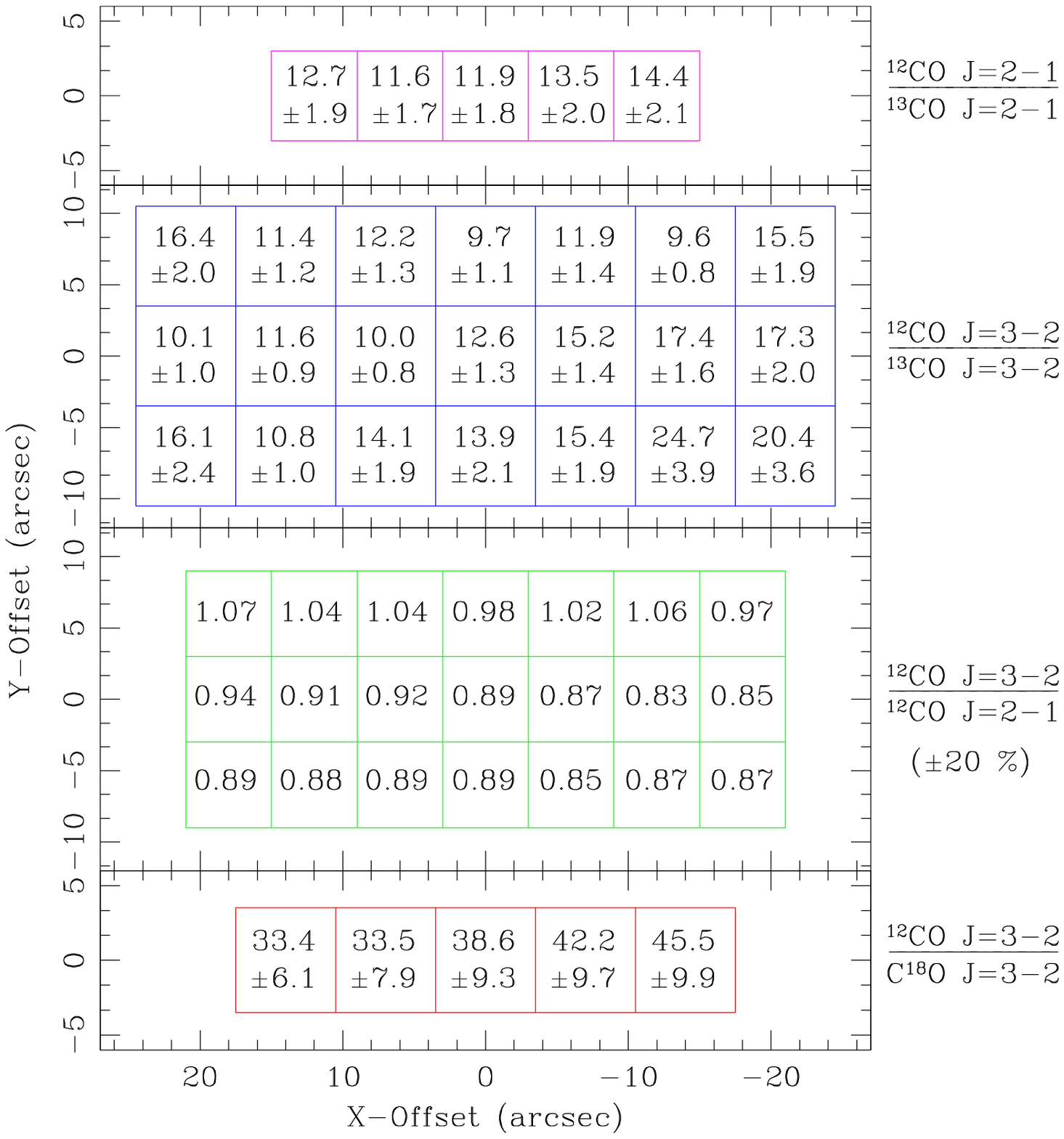}


\begin{thebibliography}{}
\bibitem[Alton, Davies, \& Bianchi(1999)]{alt99} Alton, P.~B., Davies, J.~I., Bianchi, S. 1999, \aap, 343, 51
\bibitem[Kenney \etal(1992)]{ken92} Kenney, J.~D.~P., Wilson, C.~D., Scoville, N.~Z., Devereux, N.~A., \& Young, J.~S. 1992, \apj, 395, L79
\bibitem[Kikumoto \etal(1998)]{kik98} Kikumoto, T., Taniguchi, Y., Nakai, N., Ishizuki, S., Matsushita, S. \& Kawabe, R. 1998, \pasj, 50, 309
\bibitem[Kohno, Kawabe, \& Vila-Vilar\`o(1999)]{koh99} Kohno, K., Kawabe, R., \& Vila-Vilar\`o, B. 1999, \apj, 511, 157
\bibitem[Knapp \etal(1980)]{kna80} Knapp, G.~R., Leighton, R.~B., Wannier, P.~G., Phillips, T.~G. \& Huggins, P.~J. 1980, \apj, 240, 60
\bibitem[Loiseau \etal(1990)]{loi90} Loiseau, N., Nakai, N., Sofue, Y., Wielebinski, R., Reuter, H.-P., \& Klein, U. 1990, \aap, 228, 331
\bibitem[Muxlow \etal(1994)]{mux94} Muxlow, T.~W.~B., Pedlar, A., Wilkinson, P.~N., Axon, D.~J., Sanders, E.~M. \& De Bruyn, A.~G. 1994, \mnras, 266, 455
\bibitem[Neininger \etal(1998)]{nei98} Neininger, N., Guelin, M., Klein, U., Garcia-Burillo, S. \& Wielebinski, R. 1998, \aap, 339, 737
\bibitem[Petitpas \& Wilson(1998a)]{pet98a} Petitpas, G.~R., \& Wilson, C.~D. 1998a, \apj, 496, 226
\bibitem[Petitpas \& Wilson(1998b)]{pet98b} Petitpas, G.~R., \& Wilson, C.~D. 1998b, \apj, 503, 219
\bibitem[Rickard \etal(1975)]{ric75} Rickard, L.~J., Palmer, P., Morris, M., Turner, B.~E. \& Zuckerman, B. 1975, \apjl, 199, L75
\bibitem[Seaquist, Frayer, \& Bell(1998)]{sea98} Seaquist, E.~R., Frayer, D.~T. \& Bell, M.~B. 1998, \apj, 507, 745
\bibitem[Seaquist, Frayer, \& Frail(1997)]{sea97} Seaquist, E.~R., Frayer, D.~T. \& Frail, D.~A. 1997, \apjl, 487, L131 
\bibitem[Shen \& Lo(1995)]{she95} Shen, J. \& Lo, K.~Y. 1995, \apjl, 445, L99 
\bibitem[Sutton, Masson, \& Phillips(1983)]{sut83} Sutton, E.~C., Masson, C.~R., \& Phillips, T.~G. 1983, \apjl, 275, L49 
\bibitem[Tammann \& Sandage(1968)]{tam68} Tammann, G.~A. \& Sandage, A. 1968, \apj, 151, 825
\bibitem[Tilanus \etal(1991)]{til91} Tilanus, R.~P.~J., Tacconi, L.~J., Sutton, E.~C., Zhou, S., Sanders, D.~B., Wynn-Williams, C.~G., Lo, K.~Y., \& Stephens, S.~A. 1991, ApJ, 376, 500
\bibitem[Wild \etal(1992)]{wil92} Wild, W., Harris, A.~I., Eckart, A., Genzel, R., Graf, U.~U., Jackson, J.~M., Russell, A.~P.~G. \& Stutzki, J. 1992, \aap, 265, 447

\end{thebibliography}
\end{document}